\input{aipcheck}

\documentclass[
    ,final            % use final for the camera ready runs
  ]
  {aipproc}

\layoutstyle{6x9}

\begin{document}

\title{$B_{s}^{0}$ mixing and lifetime difference measurements at CDF}


\author{Pierluigi Catastini (for the CDF Collaboration)}{
  address={Siena University and INFN sez. Pisa, Italy}
}

%\author{<author2>}{
%  address={<common address for author2 and author3>}
%}

%\author{<author3>}{
%  address={<common address for author2 and author3>}
%  ,altaddress={<author1 address>} % additional visiting address
%}

\begin{abstract}
We review latest experimental results on the $B_s$ mixing and lifetime difference measurements at CDF. We report on the latest $\beta_{s}$ and $\Delta\Gamma_{s}$ results from $B_{s}\rightarrow J/\psi \phi$. We also discuss flavor specific $\Delta\Gamma_{s}$ measurements, including information from hadronic channels, $B_{s}\rightarrow D_{s}D_{s}$ and $B_{s}\rightarrow KK$. We describe the new flavor tagging methodology and its calibration using the $B_{s}$ oscillations. 
\end{abstract}

\maketitle

%%%%%%%%%%%%%%%%%%%%%%%%%%%%%%%%%%%%%%%%%%%%
%% MAINMATTER
%%%%%%%%%%%%%%%%%%%%%%%%%%%%%%%%%%%%%%%%%%%%

%\section{Introduction}

\section{$B_{s}$ Meson Physics}
The time evolution of a mixture of the $B_{s}^{0}$ and its antiparticle $\overline{B_{s}^{0}}$ is described by the Schrodinger equation $i\frac{d}{dt}  \left( \begin{array}{c} B_{s}^{0}(t) \\ \overline{B_{s}^{0}}(t) \\  \end{array} \right) = \big ( {\boldmath M} - i\frac{{\boldmath\Gamma}}{2} \big) \left( \begin{array}{c} B_{s}^{0}(t) \\ \overline{B_{s}^{0}}(t) \\  \end{array} \right)$, where ${\boldmath M}$ and ${\boldmath \Gamma}$ are the 2$\times$2 mass and decay matrices that relate the flavor eigenstates, $B_{s}^{0}$ and $\overline{B_{s}^{0}}$, with the mass eigenstates, $B_{s}^{H}$ and $B_{s}^{L}$.  The difference in mass and width between $B_{s}^{H}$ and $B_{s}^{L}$ is related to the off diagonal elements of the mass and decay matrices as follows: $\Delta m_{s} = m_{H} - m_{L} \sim 2|M_{12}|$ and $\Delta\Gamma = \Gamma_{L} - \Gamma_{H} \sim 2|\Gamma_{12}| cos\phi_{s}$, where  $\phi_{s} = arg(-M_{12} / \Gamma_{12})$ is the CP phase and $\phi_{s} \sim 0.04$ in the standard model. The fact that the mass eigenstates are not the same as the flavour states gives rise to oscillations between the $B_{s}^{0}$ and $\overline{B_{s}^{0}}$ states with a frequency proportional to the mass difference of the mass eigenstates $\Delta m_{s}$. 
\section{Measurements of  $\Delta\Gamma_{s}$ and CP Violation Phase $\beta_s^{J/\psi \phi}$}\label{beta} 
%%%%%%%%%%%%%%% intro delta gamma
While $\Delta m_{s}$ has been measured to great precision, $\Delta\Gamma_{s}$ has so far been measured imprecisely.
To proceed with a measurement of $\Delta\Gamma_{s}$ one assumption is generally made: that the $B_{s}^{0}$ light mass eigenstate is CP
even and the heavy state is CP odd. With this assumption, two approaches to measuring $\Delta\Gamma_{s}$ are pursued. The first
is to analyse $B_{s}^{0} \rightarrow J\psi \phi$ decays, fitting the angular distributions between the decay products in order to decipher 
the CP odd and even content. The second is to measure the lifetime in a CP specific decay for which the proportion 
of CP odd and even states is known a priori. 
\subsection{Flavor Specific measurement: $B_{s}^{0} \rightarrow D_{s}^{+} D_{s}^{-}$}
\begin{figure}[!h]
\includegraphics[scale=0.27]{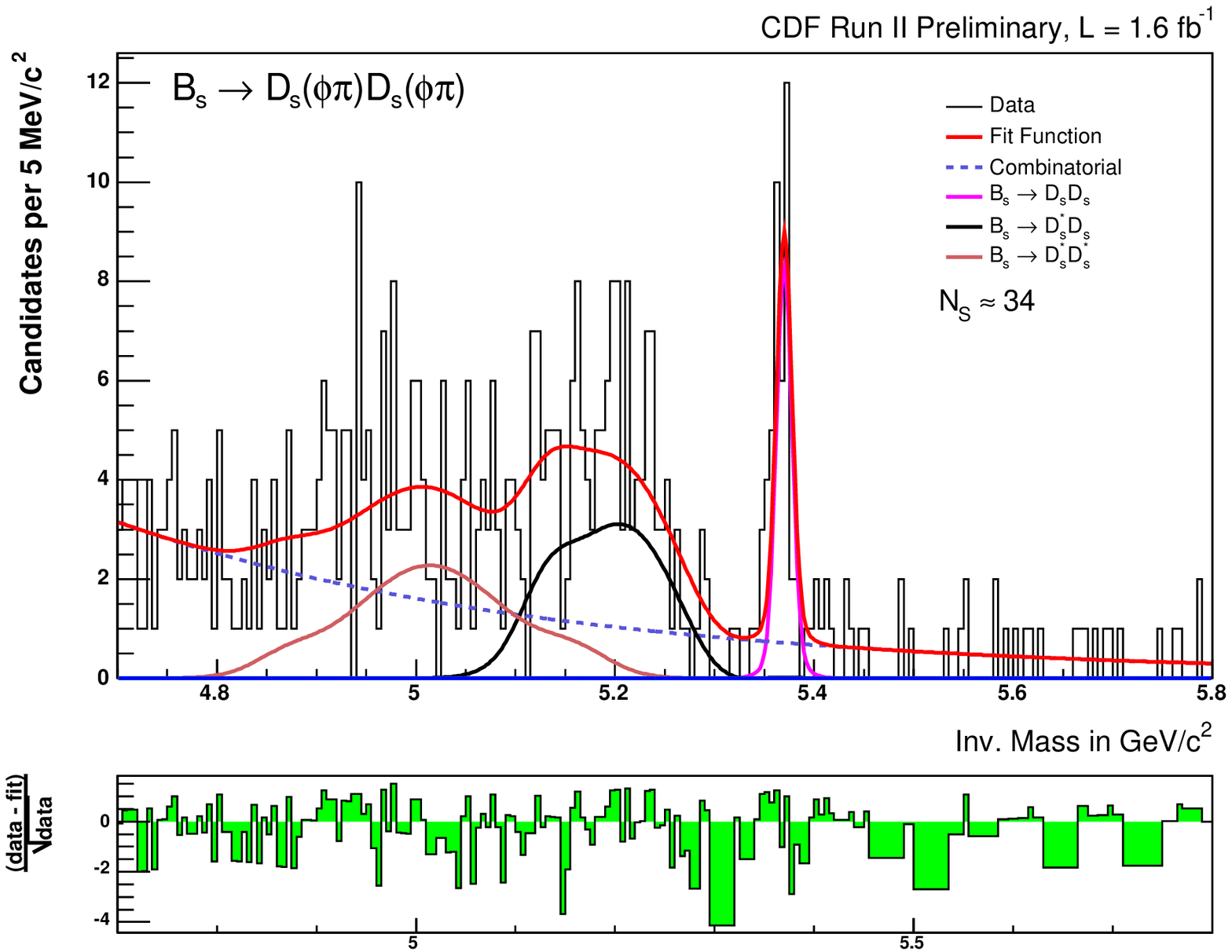}
 \includegraphics[scale=0.27]{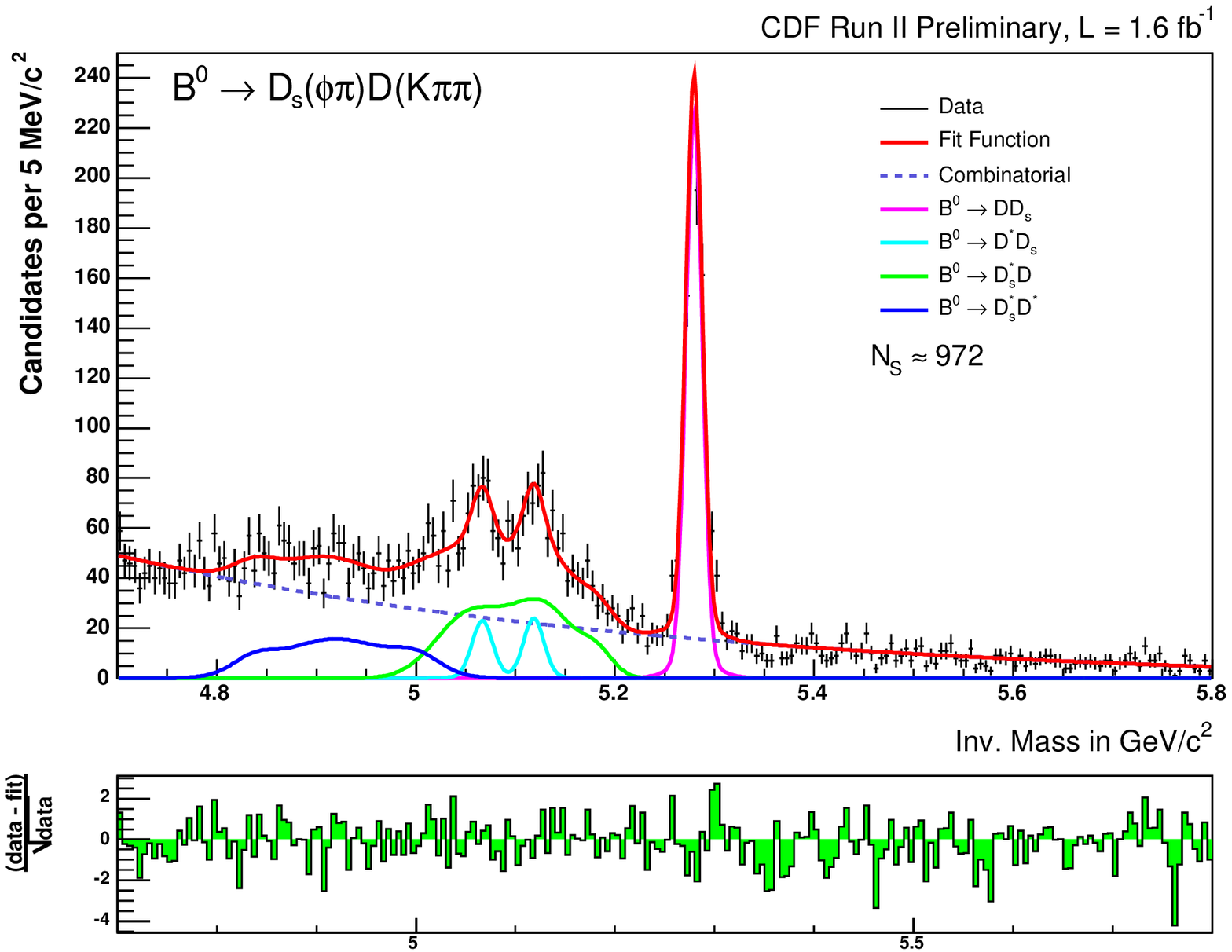}
 \caption{Fit to the invariant mass of $B_{s}^{0} \rightarrow D_{s}^{+} D_{s}^{-}$ ($D_{s}^{\pm}\rightarrow \phi \pi^{\pm}$) candidates (left) and $B^{0}\rightarrow D^{+}_{s}D^{-}$ ($D^{+}\rightarrow k\pi\pi$, $D_{s}^{\pm}\rightarrow \phi \pi^{\pm}$) candidates (right).}
 \label{dsds}
\end{figure}

%%%%%%%%%%%%%%%% DsDs
The decay $B_{s}^{0} \rightarrow D_{s}^{+} D_{s}^{-}$ is a $b\rightarrow c\bar{c}s$ decay with purely CP even composition. Therefore a lifetime measurement of
$B_{s}^{0} \rightarrow D_{s}^{+} D_{s}^{-}$ would measure $\Gamma_{L}$. Moreover, the branching ratio of the $B_{s}^{0} \rightarrow D_{s}^{+} D_{s}^{-}$ mode  provides an indirect measurement of the difference in width between the two weak eigenstates through the relation $\Delta\Gamma_{s} / \Gamma_{s} = 2 \times Br(B_{s}^{0} \rightarrow D_{s}^{+} D_{s}^{-})$ \cite{dsds}. CDF, using $355 pb^{-1}$ integrated luminosity, measured the $B_{s}^{0} \rightarrow D_{s}^{+} D_{s}^{-}$ ($D_{s}^{\pm}\rightarrow \phi \pi^{\pm}$ or $K^{*}K$ or $\pi \pi \pi$) branching ratio relative to that of $B^{0}\rightarrow D^{+}_{s}D^{-}$ 
%in order to eliminate sources of systematic uncertainty 
\cite{CDFdsds}: 
%A multi-parameter fit is performed, yielding greater than five standard deviations significance and a branching ratio of: 
%
 \begin{equation}\label{eq:dsds}
\frac{ B_{s} \rightarrow D_{s}^{+} D_{s}^{-} }{ B^{0}\rightarrow D^{+}_{s}D^{-} } = 1.44^{+0.38}_{-0.31} (stat)^{+0.08}_{-0.12} (syst)
\pm 0.21(\frac{ f_{s} }{ f_{d} }) \pm 0.20 (BR(\phi\pi)).
 \end{equation}
From this measurement, a 95\% confidence level limit of $\Delta\Gamma_{s}/\Gamma_{s} \ge 0.012$ was set.
At CDF, we are currently exploiting a new Neural Network based selection strategy to update the $\Delta\Gamma_{s}/\Gamma_{s}$ measurement in the $B_{s} ^{0}\rightarrow D_{s}^{+} D_{s}^{-}$ decay. In 1.6 $fb^{-1}$, the new selection yields $\sim 105$ $B_{s}^{0} \rightarrow D_{s}^{+} D_{s}^{-}$ and $\sim 1930$ $B^{0}\rightarrow D^{+}_{s}D^{-}$ events respectively. In Figure \ref{dsds} we show the fit to the invariant mass for the $B_{s}^{0} \rightarrow D_{s}^{+} D_{s}^{-}$ ($D_{s}^{\pm}\rightarrow \phi \pi^{\pm}$) and $B^{0}\rightarrow D^{+}_{s}D^{-}$ ($D^{+}\rightarrow k\pi\pi$, $D_{s}^{\pm}\rightarrow \phi \pi^{\pm}$) events, respectively.
%
%%%%%%%%%%%%%%%% BsKK
\subsection{Flavor Specific measurement: $B_{s}^{0}\rightarrow K^{+}K^{-} $}
The decay $B_{s}^{o}\rightarrow K^{+}K^{-} $ is $\sim 95\%$ CP even and therefore a lifetime measurement in this state can be combined with knowledge of average $B^{0}_{s}$ lifetime in order to obtain an indirect measurement of $\Delta\Gamma_{s}$. The CDF II detector collects and reconstructs significant samples of $B^{0}$ and $B_{s}^{0}$ charmless two body decays. In particular, using a data sample of $360 pb^{-1}$, CDF measured the $B_{s}^{o}\rightarrow K^{+}K^{-} $ lifetime to be $\tau(B_{s}^{0}\rightarrow K^{+}K^{-}) = 1.53 \pm 0.18 ~(stat) \pm 0.02 ~(syst) ~ps$.
 %\begin{equation}\label{eq:lifekk}
%\tau(B_{s}^{0}\rightarrow K^{+}K^{-}) = 1.53 \pm 0.18 ~(stat) \pm 0.02 ~(syst) ~ps. 
%\end{equation}
Combining the CDF measurement of $B_{s}^{0}\rightarrow K^{+}K^{-}$ lifetime with HFAG average $B_{s}^{0}$ lifetime in flavor specific decays $\tau (B_{s}^{0} ) = 1.454 \pm 0.040$ $ps$, CDF measured $\Delta\Gamma_{s} / \Delta\Gamma_{s}(B_{s}^{0}\rightarrow K^{+}K^{-}) = -0.08 \pm 0.23 ~(stat) \pm 0.03 ~(syst)$. 
%
%%%%%%%%%
%
\subsection{$\Delta\Gamma_{s}$ in $B_{s}^{0}\rightarrow J/\psi \phi $ and CP Violation Phase $\beta_s^{J/\psi \phi}$}
\begin{figure}[!h]
\includegraphics[scale=0.22]{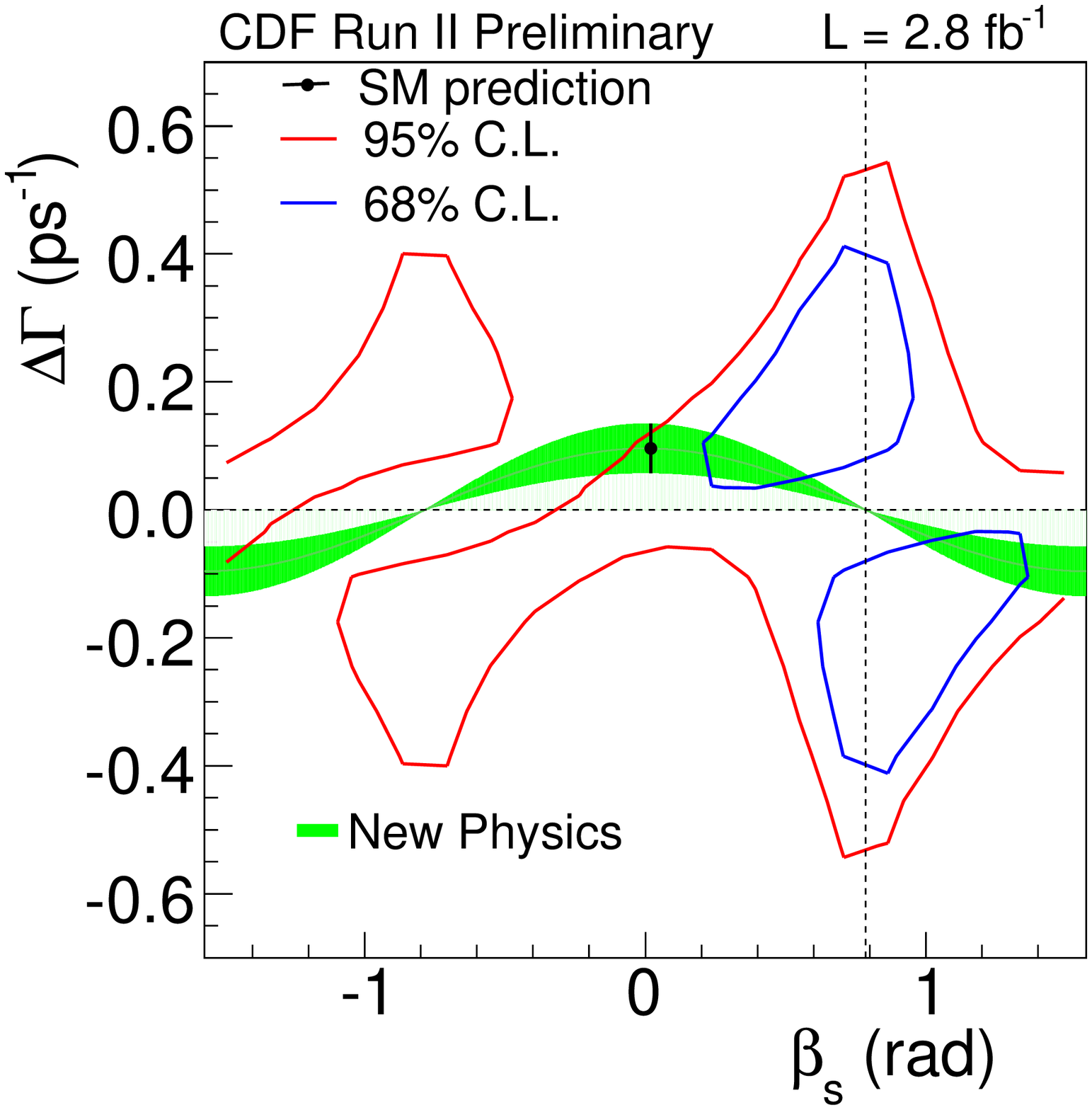}
 \includegraphics[scale=0.27]{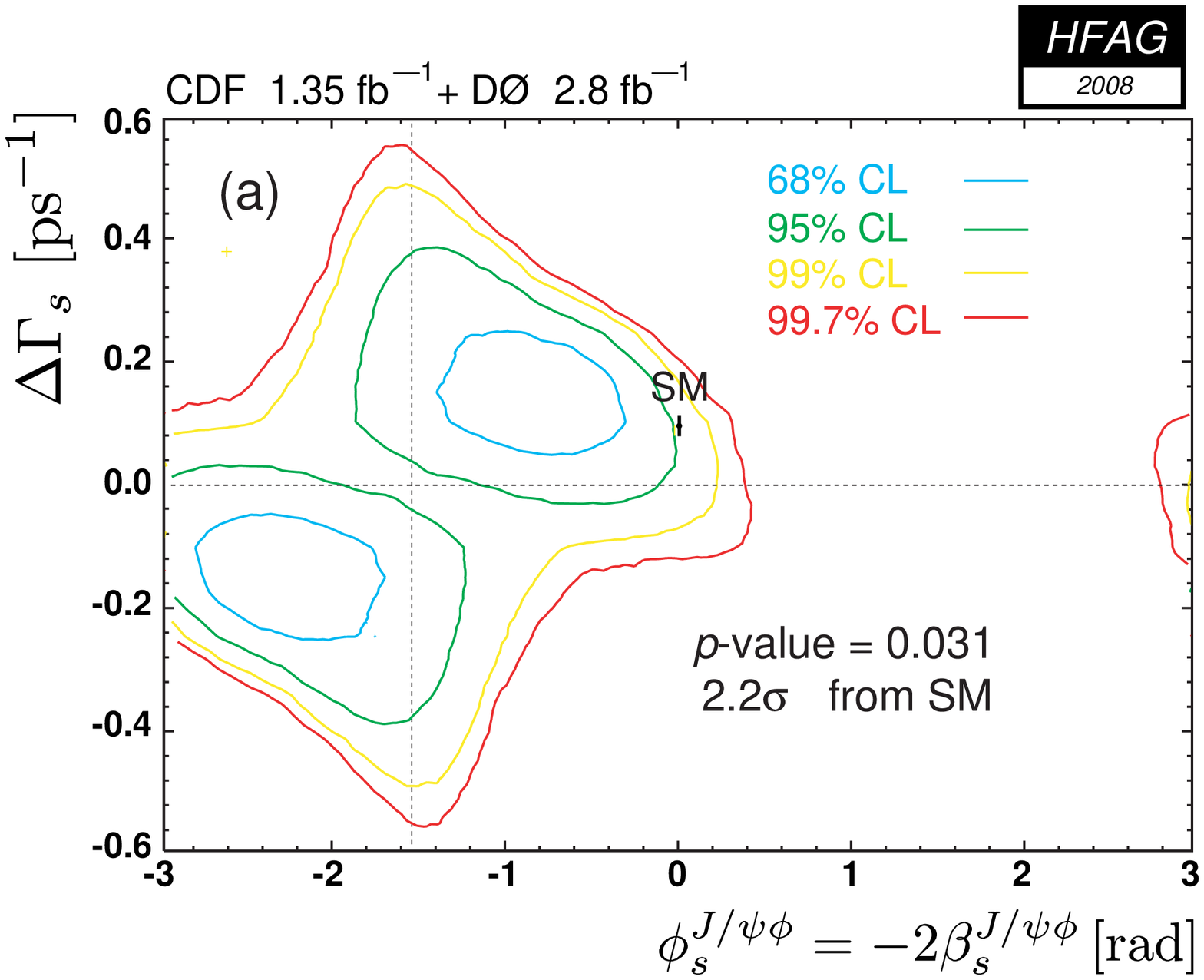}
 \caption{Confidence regions in the $\beta_s^{J/\psi \phi} - \Delta\Gamma$ plane from CDF (left). Confidence regions in the $\phi_{s}-\Delta\Gamma$ plane corresponding to the combined CDF and D0 datasets (right)}
 \label{betas_res}
\end{figure}

The study of $B_s^0\to J/\psi \phi$ decays ($J/\psi \to \mu^+ \mu^-$ and $\phi\to K^+ K^-$) allows the searching for CP non-conservation beyond the Standard Model (SM). In these decays CP violation occurs through the interference between the decay amplitudes with and without mixing. In the SM the relative phase between the decay amplitudes with and without mixing is $\beta_s^{SM}=\arg(-V_{ts}V_{tb}^*/V_{cs}V_{cb}^*)$ and it is expected to be very small~\cite{ref:SM}. New physics contributions manifested in the $B_s^0$ mixing amplitude may alter this mixing phase by a quantity $\phi_s^{NP}$ leading to an observed mixing phase $2\beta_s^{J/\psi \phi}  = 2\beta_s^{SM} - \phi_s^{NP}$. Large values of the observed $\beta_s^{J/\psi \phi}$ would be an indication of physics beyond the SM~\cite{dsds}.
The decay $B^0_s\to J/\psi \phi$ is a physics rich decay mode as it can be used to measure the $B^0_s$ lifetime, decay width difference $\Delta\Gamma_s$ and the CP violation phase $\beta_s^{J/\psi \phi}$. While the $B^0_s$ meson has spin 0, the final state $J/\psi$ and $\phi$ have spin 1. Consequently, the total angular momentum in the final state can be either 0,1 or 2. States with angular momentum 0 and 2 are CP even while the state with angular momentum 1 is CP odd. Angular distribution of the final muons and kaons from $J/\psi$ and $\phi$ decays can be used to separete the CP eigenstates. There are three angles that completely define the directions of the four particles in the final state. We use the angles $\vec{\rho}=\{\cos\theta_T, \phi_T, \cos\psi_T\}$ defined in the transversity basis introduced in~\cite{ref:dighe}.
At CDF, using 2.8 fb$^{-1}$ integrated luminosity, an unbinned maximum likelihood fit is performed to extract the parameters of interest, $\beta_s^{J/\psi \phi}$ and $\Delta\Gamma_s$.  CDF reconstructs a signal sample of $\sim 3200$ events using a Neural Network based selection. The measured $B^0_s$ lifetime and decay width difference are $\tau(B^0_s) = 1.53\pm0.04(stat)\pm0.01(syst)$ ps and $\Delta\Gamma = 0.02\pm 0.05(stat)\pm 0.01(syst)$ ps$^{-1}$. %These are the most precise measurements of the lifetime and decay width difference in the $B^0_s$ system to date. 
An exact symmetry is present in the signal probability distribution which is invariant under the simultaneous transformation ($2\beta_s\to\pi-2\beta_s$, $\Delta\Gamma\to -\Delta\Gamma$, $\delta_{||}\to 2\pi - \delta_{||}$, and $\delta_{\bot}\to pi-\delta_{\bot}$). This causes the likelihood function to have two minima. 
Confidence regions in the $\beta_s^{J/\psi \phi} - \Delta\Gamma$ plane are constructed by CDF, Figure \ref{betas_res} (left).
The resulting regions show the expected double minimum structure and are shifted with respect to the SM expectation. 
%Both results show the expected double minimum structure and they are both shifted in the same direction with respect to the SM expectation. 
%Both retsults shifted in the same direction with respect to the SM expectation. 
The significances of the deviation is 1.8 standard deviations.
%The ambiguity between the two minima could be resolved if the strong phase $\delta_{||}$ and $\delta_{\bot}$ were known. Recent theoretical studies~\cite{ref:gronau} suggest that the strong phases involved in $B_s^0\to J/\psi \phi$ decays are expected to be close to the corresponding strong phases in $B^0 \to J/\psi K^{*0}$. Using this information the CDF and D0 experiment show that in this hypothesis the preferred solution is the one corresponding to posotive $\Delta\Gamma$.
%
Combination of the CDF and D0 results has been performed~\cite{ref:hfag}. The combination includes the D0 analysis with 2.8 fb$^{-1}$~\cite{ref:d0} and a previous CDF result~\cite{ref:prevCDF} that used only 1.35 fb$^{-1}$ of data. Confidence regions, shown in Figure \ref{betas_res} (right), result in a $2.2\sigma$ deviation of $\beta_s^{J/\psi \phi}$ from the SM. Although the combined deviation from the SM expectation is not statistically significant, the independent CDF and D0 fluctuations in the same direction are interesting to follow in the future as the analyzes will be updated using more data. By the end of the Tevatron running, samples of 8 fb$^{-1}$ are expected. 
A joint effort of the CDF and D0 is currently underway to produce Tevatron combined confidence regions in the $\Delta\Gamma_{s} - \beta_{s}$ plan. Two approaches are pursued: the first combines the bi-dimensional profile likelihoods of the two experiments, a preliminary result can be found in \cite{newcomb}; the second approach, more powerful but on a longer time scale, will perform a simultaneous fit of the CDF and D0 data.
%
%Figure \ref{comb} shows the CDF probability of observing a $5\sigma$ deviation from the SM as a function of $\beta_s^{J/\psi \phi}$ assuming $\Delta\Gamma = 0.1$ ps$^{-1}$. The extrapolation assumes no further improvements of the analysis. However, improvements in the use of particle identification, tagging power and sample size by using additional triggers are expected from CDF while D0 will optimize the signal selection for better signal to background.
%
\section{Measurement of $B_{s}$ mixing and Flavor Tagging}\label{mixing}
CDF observed $B_{s}^{0}$ mixing and measured $\Delta m_{s} = 17.77 \pm 0.12 ~ps^{-1}$ with remarkable precision \cite{bsmix}. %$B_{s}$ mixing is now a very important benchmark for flavor tagging calibration. In fact, precise knowledge of 
Good sensitivity to $\Delta m_{s}$ can be exploited to calibrate improved flavor tagging algorithms. At CDF, a new flavor tagging algorithm is under development. The new approach combines the information of alla the tracks in the event. For a given $B_{s}$ candidate, tracks are divided in three categories: 1) tracks in the same side in wich the $B_{s}$ was found; 2) tracks that are an electron or muon candidate; 3) all remaining tracks. For each category, a track flavor correlation neural network is trained; finally, the output of the track flavor correlation neural networks are combined in a likelihood ratio. The new flavor tagger is currently calibrated and checked on Monte Carlo samples and on a new $B_{s}^{0}$ mixing measurement.
The new flavor tagger will be used in  the flavor tagged CP Violation Phase $\beta_s^{J/\psi \phi}$ analysis.
%
%\begin{figure}[!h]
%\includegraphics[scale=0.22]{2d_contours_CDF.eps}
% %\includegraphics[scale=0.27]{2d_contours_comb.eps}
% \caption{Comparison of Dilution for the old and new tagger in fully reconstructed $B^{+}$ events.}
% \label{tagging}
%\end{figure}
%
%
\bibliographystyle{aipproc}   % if natbib is available
%\bibliographystyle{aipprocl} % if natbib is missing

%%%%%%%%%%%%%%%%%%%%%%%%%%%%%%%%%%%%%%%%%%%
%% You probably want to use your own bibtex database here
%%%%%%%%%%%%%%%%%%%%%%%%%%%%%%%%%%%%%%%%%%%
\bibliography{sample}

\begin{thebibliography}{99}

\bibitem{dsds} 
I.~Dunietz et al. \emph{Phys.Rev. D} \textbf{63}, 114015 (2001).

\bibitem{CDFdsds}
T.~Aaltonen et al. (CDF Collaboration), \emph{Phys. Rev. Lett.} \textbf{100}, 021803 (2008).
\bibitem{ref:SM} 
I.~I.~Y.~Bigi, A.~I.~Sanda, \emph{Nucl. Phys.} \textbf{B193}, 85 (1981).  

\bibitem{ref:dighe}
 A.~S.~Dighe, I.~Dunietz, R.~Fleischerde,  \emph{Eur. Phys. Rev. J. C} \textbf{6}, 647 (1999). 

\bibitem{ref:hfag} 
E.~Barberio et al., \emph{Heavy Flavor Averaging Group}, 2007 http://www.slac.stanford.edu/xorg/hfag.

\bibitem{ref:d0} 
V.~Abazov et al. (D0 Collaboration),  \emph{Phys. Rev. Lett.} \textbf{101}, 241801 (2008).

\bibitem{ref:prevCDF}
T.~Aaltonen et al. (CDF Collaboration), \emph{Phys. Rev. Lett.} \textbf{100}, 121802 (2008).

\bibitem{newcomb}
%http://www-cdf.fnal.gov/physics/new/bottom/090721.blessed-betas\_combination2.8/D0Note5928\_CDFNote9787.pdf
http://www-cdf.fnal.gov/physics/new/bottom/090721.blessed-betas\_combination2.8/

\bibitem{bsmix}
A.~Abulencia et al. (CDF Collaboration), \emph{ Phys.Rev.Lett.} \textbf{97}, 062003 (2006).

%    \bibitem{ref:babar1} \BY{Aubert~B. et al. (BaBar Collaboration).}
%  \IN{Phys. Rev. D}{77}{2008}{111102}.
% \bibitem{ref:belle1} \BY{Abe~K. et al. (Belle Collaboration).}
%  \IN{Phys. Rev. D}{73}{2006}{051106}.
%   \bibitem{ref:bsdsk} \BY{Abulencia~A. et al. (CDF Collaboration).}
%  arXiv:0809.0080v1[hep-ex].
%  \bibitem{ref:SM} \BY{Bigi~I.~I.~Y. \atque Sanda~A.~I.}
%  \IN{Nucl. Phys.}{B193}{1981}{85}.  
%   \bibitem{ref:beyondSM} \BY{Dunietz~I, Fleisher~R., \atque Nierste~U.}
%  \IN{Phys. Rev. D}{63}{2001}{114015}.  
%\bibitem{ref:dighe} \BY{Dighe~A.~S., Dunietz~I. \atque Fleischerde~R.}
%  \IN{Eur. Phys. Rev. J. C}{6}{1999}{647}.  
%  \bibitem{ref:babar} \BY{Aubert~B. et al. (BaBar Collaboration).}
%  \IN{Phys. Rev. D}{76}{2007}{031102(R)}.
% \bibitem{ref:belle} \BY{Itoh~R. et al. (Belle Collaboration).}
%  \IN{Phys. Rev. Lett.}{95}{2005}{091601}.
%  \bibitem{ref:cdf} http://www-cdf.fnal.gov/physics/new/bottom/080724.blessed$-$tagged$\_$BsJPsiPhi$\_$update$\_$prelim
%     \bibitem{ref:d0} \BY{Abazov~V. et al. (D0 Collaboration).}
%  \IN{Phys. Rev. Lett.}{101}{2008}{241801}.
%  %\bibitem{ref:gronau} \BY{Gronau~M.  \atque Rosner~J.~L.}
%  %\IN{Phys. Lett. B}{669}{2008}{321-326}.  
%  \bibitem{ref:hfag} \BY{Barberio~E. et al.}, Heavy Flavor Averaging Group, 2007 http://www.slac.stanford.edu/xorg/hfag.
%   \bibitem{ref:prevCDF} \BY{Aaltonen~T. et al. (CDF Collaboration).}
%  \IN{Phys. Rev. Lett.}{100}{2008}{121802}.



\end{thebibliography}

%%%%%%%%%%%%%%%%%%%%%%%%%%%%%%%%%%%%%%%%%%%
%% Just a reminder that you may have to run bibtex
%% All of it up to \end{document} can be removed
%% if you don't like the warning.
%%%%%%%%%%%%%%%%%%%%%%%%%%%%%%%%%%%%%%%%%%%
\IfFileExists{\jobname.bbl}{}
 {\typeout{}
  \typeout{******************************************}
  \typeout{** Please run "bibtex \jobname" to optain}
  \typeout{** the bibliography and then re-run LaTeX}
  \typeout{** twice to fix the references!}
  \typeout{******************************************}
  \typeout{}
 }

%%%%%%%%%%%%%%%%%%%%%%%%%%%%%%%%%%%%%%%%%%%
%% The following lines show an example how to produce a bibliography
%% without the help of the BibTeX program. This could be used instead
%% of the above.
%%%%%%%%%%%%%%%%%%%%%%%%%%%%%%%%%%%%%%%%%%%

\end{document}